# Generation and dynamics of solitonic pulses due to pump amplitude modulation at normal group velocity dispersion

Valery E. Lobanov[1], Nikita M. Kondratiev[1], Artem E. Shitikov[1,2], Ramzil R. Galiev[1,2], Igor A. Bilenko[1,2]

*[1]Russian Quantum Center, Skolkovo 143025, Russia*

*[2]Faculty of Physics, Lomonosov Moscow State University, Moscow 119991, Russia*

## Abstract

We studied generation and dynamics of solitonic pulses, platicons, at normal group velocity dispersion due to the pump amplitude modulation. We proposed, that if the required frequency of amplitude modulation is too large for available modulators, it is possible to use subharmonic phase modulation for platicon generation. It was also demonstrated that it is possible to control the repetition rate of platicons by tuning modulation frequency. Tuning range for platicons was found to be much wider than for bright solitons. We also studied the influence of the high-order dispersion on platicon properties. It was shown that both third-order dispersion and pump modulation govern platicon dynamics making it quite different from dynamics of dissipative Kerr solitons. Third-order dispersion was also shown to affect significantly the optimal conditions for the platicon generation and repetition rate tuning range.

## 1. Introduction.

In recent years, soliton Kerr frequency combs [1, 2] generated in high-quality-factor (high-Q) microresonators have proven to be unique tools for various fields of science and technology, including spectroscopy [3-5], astrophysical measurements [6, 7], LIDARs [8], low-noise microwave generation [9] and telecommunication systems [10, 11]. However, the area of application of Kerr frequency combs is often limited by the spectral ranges characterized by anomalous group velocity dispersion (GVD) since achieving modulation instability for comb initiation at normal GVD is a challenging task [12, 13]. At the same time, the material GVD of microresonator is usually normal in visible or telecommunication frequency range. One may achieve anomalous GVD even in such spectral ranges by engineering the resonator dispersion via the resonator geometry [14-16], however, such a process may be rather tricky. Nevertheless, mode-locked Kerr frequency combs in the normal dispersion regime were experimentally demonstrated in different settings [17, 18]. It was shown numerically that in some cases such experimental results may be explained using a novel type of solitonic pulses called "platicons," flat-topped bright pulses that can be softly excited and stably exist in microresonators with normal dispersion under the condition of local dispersion perturbation, e.g. pump mode shift [19]. In real microresonators, this condition can be fulfilled due to the normal mode coupling between different mode families [20, 21] or, presumably, due to the self-injection locking effect [17, 22]. Platicons may be interpreted as bound states of opposing switching waves in microresonator that connect upper and lower branches of bistable nonlinear resonance to satisfy periodic boundary conditions [23, 24]. Taking the spatio-temporal analogy into account one may notice that similar scenario of the formation of positive and negative autosolitons due to the diffractive coupling of the switching waves was demonstrated in wide-aperture driven nonlinear cavities [25]. Further, it was shown for platicons that one may control their duration in a wide range varying the pump detuning. Generation of platicons was found to be significantly more efficient than the generation of bright soliton trains in microresonators in terms of conversion of the cw pump power into the power of the comb [26]. Conversion efficiency exceeding 30% was demonstrated experimentally in the fiber telecom band by employing dark pulse mode-locking in the normal dispersion range while the conversion efficiency of bright solitons is generally limited to a few percent [27, 28]. In [29], it was demonstrated numerically that the dynamics of platicons in the presence of the third-order dispersion is quite peculiar and drastically different from bright solitons dynamics [30]. In [31], a possibility of stable coexistence of dark and bright solitons in case of nonzero third-order dispersion was revealed. In [32] it was shown that Raman scattering may induce instability of the platicon pulses resulting in branching of platicons and complex spatiotemporal dynamics. Interestingly, platicon generation is also possible in absence of the local dispersion perturbation when a bichromatic or an amplitude-modulated pump is used [33]. This method is efficient if pump modulation frequency or frequency difference between two pump waves is equal to a free spectral range (FSR, the inverted round-trip time of light in the microresonator) of the microresonator. The feasibility of this method was confirmed experimentally [34]. Such an approach seems to be the most simple for the experimental realization.

In this work, we report results of numerical analysis of the process of platicon generation via an amplitude-modulated (AM) pump and demonstrate new features of platicon generation and nontrivial platicon dynamics that is important for understanding of complex dynamics of localized dissipative structures and useful for future experiments. We propose the possibility of application of the pump phase modulation (PM) at lower frequencies. We also estimate the platicon repetition rate tuning range in comparison with the tuning range of bright solitons and show that it may be wider significantly. It is revealed that platicon dynamics is governed by both amplitude modulation and third-order dispersion (TOD) and it may be quite different from dynamics of dissipative Kerr solitons at anomalous GVD and dynamics of platicons generated due to pump mode shift. TOD is also shown to affect optimal generation conditions leading to a change of the required modulation frequency.

## 2. Subharmonic modulation.

Using amplitude-modulated pump for platicon generation one may encounter the problem that the microresonator FSR (from tens of GHz to THz) may exceed substantially the maximum modulation frequency of the available amplitude modulators (usually less than 20 GHz). In recent works studying different ways for the single-soliton generation [35, 36], a method to solve this problem by using phase modulation at fractional frequencies (subharmonics of the FSR) was proposed. Note, that resonant phase modulation with modulation frequency equal to microresonator FSR does not provide platicon generation [33]. However, one may use strong PM at even subharmonics of microresonator FSR to obtain effective amplitude modulation FSR. Using well-known expansion $e^{i\varepsilon\sin\Omega t} = \sum_m J_m(\varepsilon)e^{im\Omega t}$, where $\varepsilon$ is PM depth, $J_m(\varepsilon)$ is the Bessel function of the order $m$, and taking into account that $J_{-m}(\varepsilon) = (-1)^m J_m(\varepsilon)$, it may be shown that at modulation frequency $\Omega$ equal to FSR/$p$ PM mimics AM for resonant spectral modulation components, if $p=2k$ ($p$ is a subharmonic number, $k$ is an integer). Thus, one may obtain a simple formula for the effective AM depth: $\varepsilon_{eff} = 2J_p(\varepsilon)/J_0(\varepsilon)$ [see Fig. 1]. Using this expression one may show that to obtain AM of 0.3 that is quite enough for platicon generation one may use phase modulation at the second subharmonic ($p=2$) with the modulation depth of 1.0 ($\varepsilon = 1.12$ for $\varepsilon_{eff} = 0.4$). However, since the growth of PM depth leads to decrease of primary mode pump power, pump power should be increased by the factor $K_p = (1/J_0(\varepsilon))^2$ that is $K_2 = 1.71$ for $\varepsilon = 1.0$ and $K_2 = 1.98$ for $\varepsilon = 1.12$. The feasibility of this method was checked numerically using coupled mode approach described in [33,35,37]. We also checked that in most cases under such conditions it is enough to consider first resonant sidebands in the modulation spectrum leaving only three terms in the modulation function expansion since other resonant spectral components are negligible and do not affect platicon generation process.

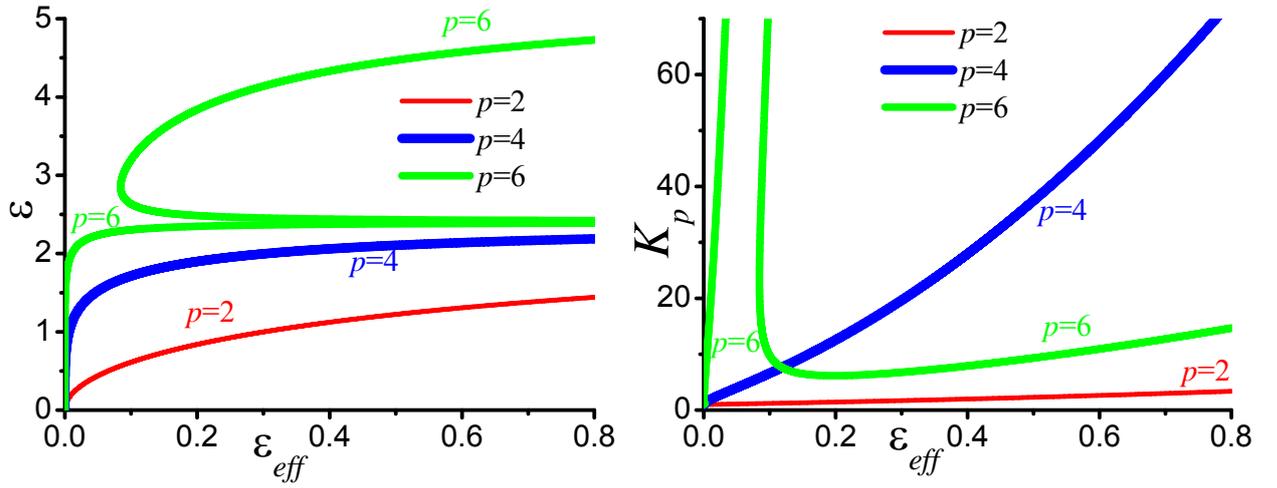

**Fig. 1.** (left panel) $\varepsilon$ vs. $\varepsilon_{eff}$ and (right panel) $K_p$ vs. $\varepsilon_{eff}$ for different subharmomic numbers $p$.

Proposed approach allows to generate coherent frequency combs or platicons in high-Q crystalline microresonators at normal GVD using accessible commercial phase modulators.

## 3. Third-order dispersion.

We also checked numerically the influence of the third-order dispersion on platicon generation using both coupled mode approach [33, 35] and Lugiato–Lefever equation (LLE) [29, 38, 39]. The influence of TOD is well studied for generation of dissipative Kerr solitons at anomalous GVD [30, 40-45] but insufficiently for dark solitons and platicons. Dispersive wave emission from dark solitons was studied in [46]. Also, it was found that platicons may be generated in the case of the pumped mode shift even if the TOD dispersion is present [29]. Platicon generation and existence domains depend weakly on the TOD coefficient value. Also, due to the influence of TOD, the platicons obtain drift velocity depending both on dispersion and on detuning value.

Nonlinear coupled mode equations were modified to take the pump modulation $f(t) = F(1+\varepsilon \cos \Omega t)$ into account [33, 35]:

$$\frac{\partial a_\mu}{\partial \tau} = -(1+i\zeta_\mu)a_\mu + i\sum_{\mu' \leq \mu''}(2-\delta_{\mu'\mu''})a_{\mu'}a_{\mu''}a^*_{\mu'+\mu''-\mu} + f_\mu \exp(i\mu\Delta\tau). \qquad (1)$$

We consider Taylor expansion of the dispersion law $\omega_\mu = \omega_0 + D_1\mu - \frac{1}{2}D_2\mu^2 + \frac{1}{6}D_3\mu^3$, where $\omega_\mu$ are microresonator eigenfrequencies, $\omega_0$ corresponds to the pumped mode, $D_1 = 2\pi / T_R$ is the FSR of the microresonator, $T_R$ is the round-trip time, $D_2$ and $D_3$ are GVD and TOD coefficients, correspondingly. All mode numbers $\mu$ are defined relative to the pumped mode $\mu = m - m_0$ with the initial azimuthal number $m_0 \approx 2\pi R n_0 / \lambda$, where $\lambda = 2\pi c / \omega_0$ is the wavelength. Note, that there is a minus before the quadratic term since GVD is normal (we assume that $D_2 > 0$). Here $\zeta_\mu = 2(\omega_\mu - \omega_p - \mu D_1)/\kappa$ is the normalized detuning, $\omega_p$ is pump frequency, $a_\mu$ is the slowly varying amplitude of the comb modes for the mode frequency $\omega_\mu$, $\tau = \kappa t / 2$ denotes the normalized time, $\kappa = \omega_0 / Q$ denotes the cavity decay rate, $Q$ is the total quality factor. $f_{-1,0,1} = F\{\varepsilon/2, 1, \varepsilon/2\}$, $F = \sqrt{\dfrac{8g\eta P_0}{\kappa^2 \hbar \omega_0}}$ stands for the dimensionless pump amplitude, $g = \dfrac{\hbar \omega_0^2 c n_2}{n_0^2 V_{eff}}$ is the nonlinear coupling coefficient, $V_{eff}$ is the effective mode volume, $n_2$ is the nonlinear refractive index, $\eta$ is the coupling efficiency ($\eta = 1/2$ for the critical coupling). $\varepsilon$ is the modulation depth, $\Delta = 2(D_1 - \Omega)/\kappa$ is the normalized modulation frequency mismatch. The corresponding LLE was the following [29]:

$$\frac{\partial \psi}{\partial \tau} = -i\frac{\beta_2}{2}\frac{\partial^2 \psi}{\partial \varphi^2} + \frac{\beta_3}{6}\frac{\partial^3 \psi}{\partial \varphi^3} + i|\psi|^2 \psi - (1+i\zeta_0)\psi + F(1+\varepsilon\cos(\varphi + \Delta\tau)), \quad (2)$$

where $\varphi \in [-\pi; \pi]$ is an azimuthal angle in a coordinate system rotating with the angular frequency equal to $D_1$, $\psi(\varphi) = \sum_\mu a_\mu \exp(i\mu\varphi)$ is the slowly varying waveform describing field azimuthal distribution inside the microresonator, $\beta_2 = 2D_2/\kappa$, $\beta_3 = 2D_3/\kappa$.

At first, we studied numerically the generation of platicons from a noise-like input by the frequency scan ($\zeta_0 = \zeta_0(0) + \alpha\tau$) for different values of the third-order dispersion coefficient $\beta_3$. Eq. 2 was solved numerically using standard split-step Fourier routine with 1024 points in the azimuthal direction. To check simulation results, coupled-mode equations (1) for 1024 modes were numerically propagated in time using the adaptive Runge–Kutta integrator. Nonlinear terms were calculated using a fast method proposed in [47]. We also checked that results do not change with the increase of number of modes. Results obtained by two methods were found to be in a good agreement.

We set $\Delta = 0$, $F = 4$, $\beta_2 = 0.005$, $\alpha = 0.002$ and studied field distribution evolution upon frequency scan (see Fig. 1). It should be noted that qualitatively similar results were also obtained for different values of these parameters. First, we observed platicon generation from the noise-like input in the absence of the third-order dispersion. In Fig. 1(a) one may notice that at some detuning value abrupt change of the field distribution takes place indicating platicon formation. To be confident that steady-state solutions were reached we repeated simulation with significantly smaller frequency scan velocity $\alpha$. We also checked that in the absence of the frequency scan generated patterns propagate in a stable fashion over indefinitely large periods of time.

Then we studied the influence of the third-order dispersion and it was found out that it affects platicon generation process greatly. It was revealed that in contrast to the case of dissipative Kerr solitons at anomalous GVD [30] and platicon generation due to pump mode shift described in [29], platicons generated via AM pump does not experience a drift upon TOD. Such platicons are localized in the vicinity of pump maximum at $\varphi = 0$ while TOD tries to shift them [see Figs. 2(a) and 2(b)] making field distribution asymmetric.

If TOD value is large enough platicons begin to move away from pump maximum and decays rapidly [see Figs. 2(c) and 2(d)]. Platicon generation frequency range (range of pump detuning $\zeta_0$) decreases notably with the growth of

TOD value. Above some critical TOD value generation almost disappears and in order to avoid this larger modulation depth should be used.

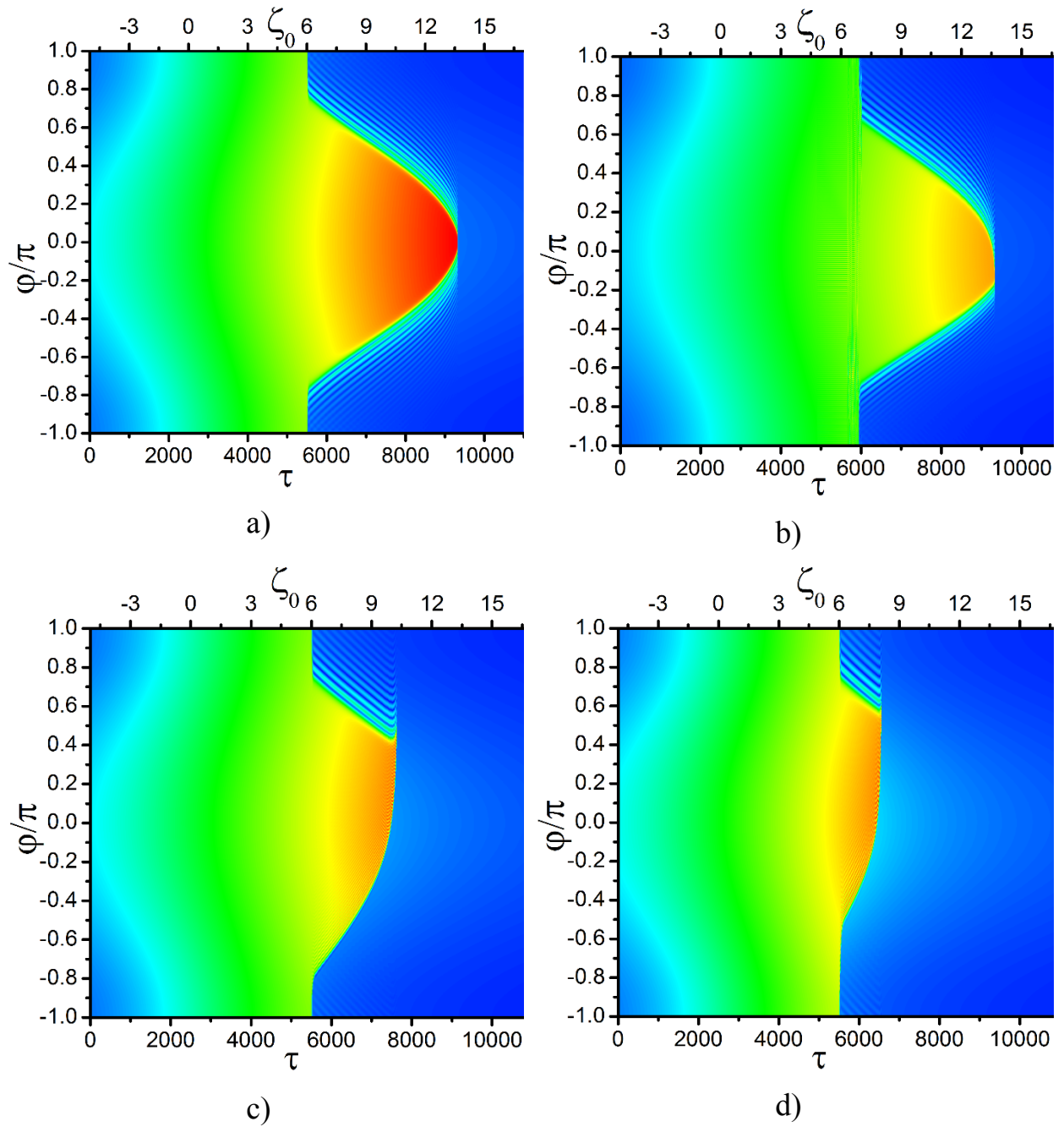

**Fig. 2.** Field distribution evolution upon frequency scan ($\zeta_0 = \zeta_0 + \alpha\tau$, $\alpha = 0.002$) in the temporal representation at AM depth $\varepsilon = 0.4$ and $\beta_2 = 0.005$ for different values of the TOD coefficient: a) $\beta_3 = 0$, b) $\beta_3 / 3\beta_2 = 0.004$, c) $\beta_3 / 3\beta_2 = 0.012$, d) $\beta_3 / 3\beta_2 = 0.02$.

To reveal the influence of the TOD on platicon dynamics, we simulate platicon propagation at fixed mismatch value for different values of $\beta_3$ using platicon profiles obtained for $\beta_3 = 0$ as an input. As it is shown in Fig. 3, platicons propagate in a stable manner without drift up to the critical TOD value. If $\beta_3$ exceeds this critical value platicon drift and decay is observed [see Fig. 3(f)]. In Fig. 3(c) one may observe a scenario of counteraction of the pump modulation to TOD: drift tries to take platicon away while pump gradient returns it back. Note, that direction of TOD action (drift direction) depends on the TOD value [compare Figs. 3(b) and 3(e), 3(a)] that is in a good agreement with results reported in [29].

Also, with the growth of $\beta_3$, the platicon profile (flat-top with symmetric oscillating tails) becomes indented and asymmetric with a pronounced one-sided oscillating tail [see Fig. 4]. Platicon spectrum also becomes asymmetric and even wider in the presence of TOD.

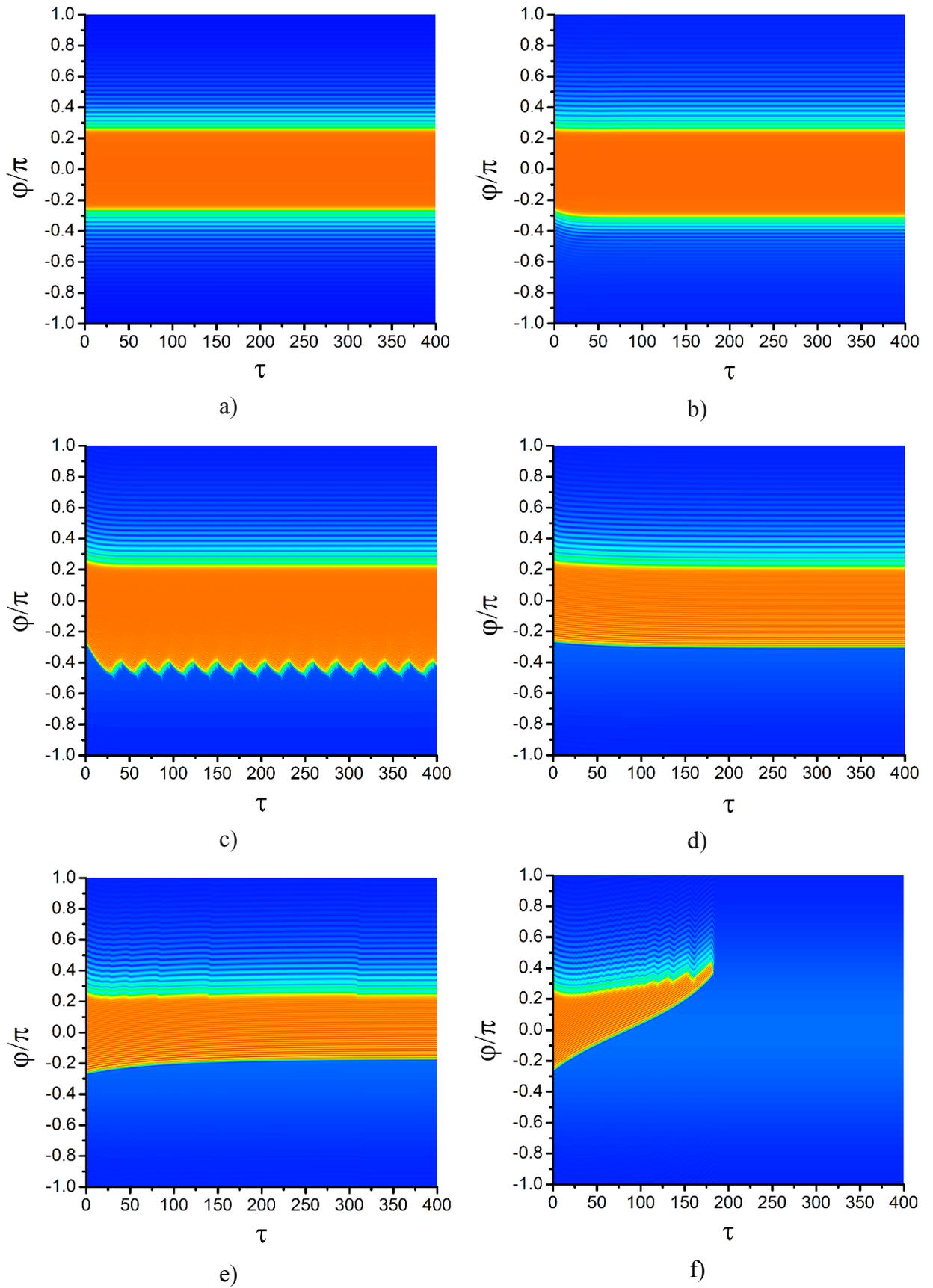

**Fig. 3.** Platicon propagation at $\zeta_0 = 13$, $\beta_2 = 0.005$ and $\varepsilon = 0.5$ for different values of the TOD: (a) $\beta_3 / 3\beta_2 = 0.0$; (b) $\beta_3 / 3\beta_2 = 0.004$; (c) $\beta_3 / 3\beta_2 = 0.006$; (d) $\beta_3 / 3\beta_2 = 0.008$, (e) $\beta_3 / 3\beta_2 = 0.0084$, (f) $\beta_3 / 3\beta_2 = 0.0092$.

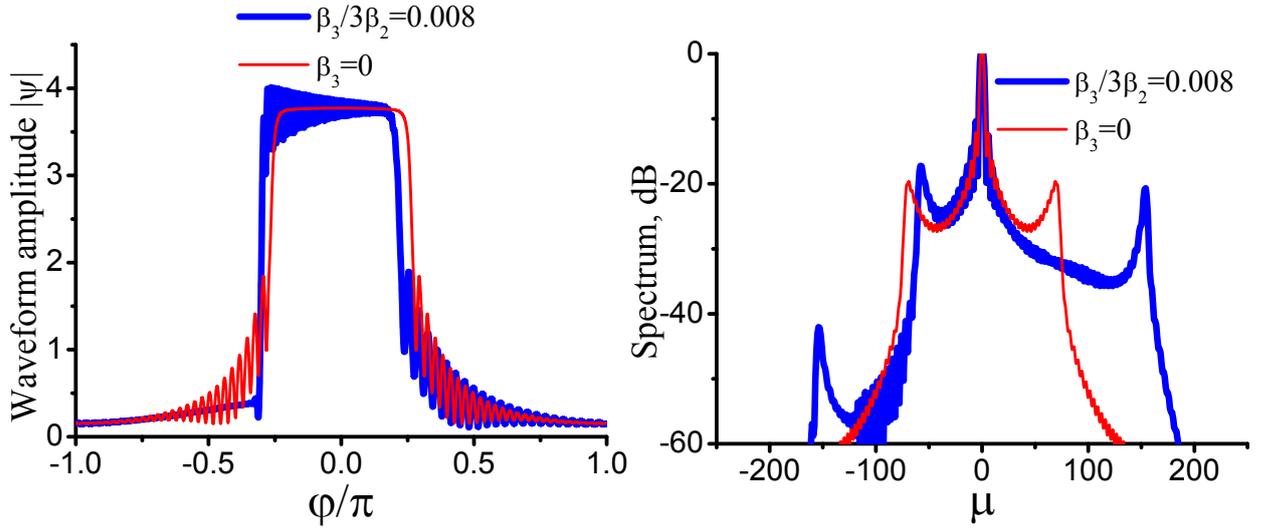

**Fig. 4.** Platicon profiles in dimensionless units (left panel) and spectra (right panel) at $\zeta_0 = 13$, $\beta_2 = 0.005$, $\varepsilon = 0.5$ for different values of the third-order dispersion.

The critical value of TOD $\beta_{3cr}$ decreases with the growth of the pump detuning value $\zeta_0$ [see Fig. 5, left panel]. For example, at $\varepsilon = 0.5$, $\beta_2 = 0.005$ platicon decays if $\beta_3/3 > 0.000042$ ($\beta_3/3\beta_2 > 0.0084$) for $\zeta_0 = 13$ and if $\beta_3/3 > 0.0001$ ($\beta_3/3\beta_2 > 0.02$) for $\zeta_0 = 8$. Thus, narrow platicons are more sensitive to the TOD than wide platicons. This result agrees with the fact reported in [29] that in the presence of the pump mode shift the absolute value of platicon drift velocity (or, in other words, TOD influence) increases with the growth of $\zeta_0$. Thus, for the same value of $\beta_3$, the force trying to shift platicon from pump maximum increases with the growth of $\zeta_0$.

Critical TOD value also increases with the growth of GVD coefficient $\beta_2$. Considering the normalization procedure for the Eq. 2, one may approximate this dependence as $\beta_{3cr} \approx const * \beta_2^{3/2}$ and one may see a good correspondence between this estimation and numerical results in Fig. 5.

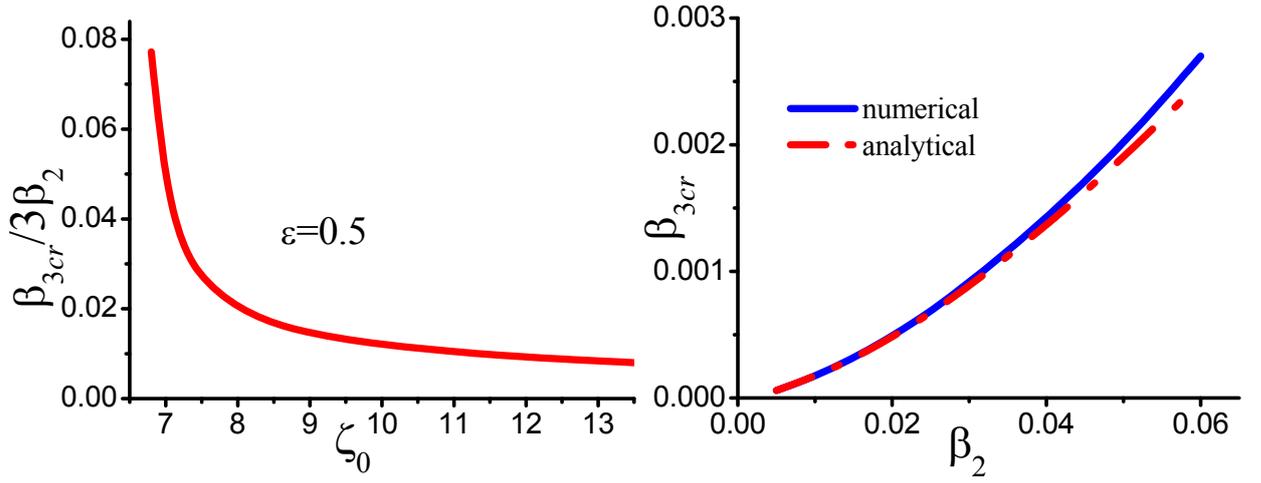

**Fig. 5.** (left panel) Critical TOD value $\beta_{3cr}$ vs. pump frequency detuning $\zeta_0$ at $\beta_2 = 0.005$, $\varepsilon = 0.5$. Platicons exist for the parameters below the red line. (right panel) $\beta_{3cr}$ vs. $\beta_2$ at $\zeta_0 = 10$, $\varepsilon = 0.5$. All quantities are plotted in dimensionless units.

### 4. Platicon injection locking.

It is well-known that one may control the repetition rate of dissipative Kerr solitons generated in optical microresonators [1,2] via microwave injection-locking technique [48-51]. The injection locking of the soliton repetition rate may be implemented by the applying amplitude modulation or phase modulation on the pump laser, at a frequency close to the FSR. In this case, a modulated cw field traps solitons regulating time interval between them. Thus, in the frequency domain modulation frequency defines soliton repetition rate.

Here we studied the possibility of platicon repetition rate control by tuning the frequency of a pump amplitude modulation. It should be noted that while bright dissipative Kerr solitons can be generated without pump modulation that may be used additionally for a single-soliton generation [35], AM is a crucial aspect for a platicon generation. Eq. 2 was rewritten in a coordinate system rotating with the modulation frequency using the substitution $\theta = \varphi + \Delta\tau$:

$$\frac{\partial \psi}{\partial \tau} + \Delta \frac{\partial \psi}{\partial \theta} = -i\frac{1}{2}\beta_2 \frac{\partial^2 \psi}{\partial \theta^2} + i|\psi|^2 \psi - (1+i\zeta_0)\psi + F\{1+\varepsilon\cos\theta\}. \quad (3)$$

Note, that Eq. 3 contains drift term which strength is proportional to the modulation frequency mismatch value. Then we searched for the stationary solutions of Eq. 3:

$$-\frac{1}{2}\beta_2 \frac{\partial^2 \psi}{\partial \theta^2} + i\Delta \frac{\partial \psi}{\partial \theta} + |\psi|^2 \psi + i(1+i\zeta_0)\psi - iF\{1+\varepsilon \cos\theta\} = 0. \qquad (4)$$

Such solutions describe platicons which repetition rate is equal to the modulation frequency. Eq. 4 was solved numerically by means of the relaxation technique. We set $F = 4$, $\beta_2 = 0.02$. First of all, we calculated the dependence of platicon energy $U = \int_{-\pi}^{\pi} |\psi|^2 d\varphi$ on detuning value $\zeta_0$ and found platicon existence domains for different values of the modulation depth $\varepsilon$ [see Fig. 6, left panel]. While the modulation depth increases, the existence domain becomes wider and the energy spectrum contains a smaller number of steps. Discrete energy levels correspond to wide platicons with different number of oscillations in platicon profile [19,33].

Interestingly, even for high values of the modulation depth symmetry breaking effect reported for bright solitons [52] was not found. It was revealed that for each value of the modulation depth $\varepsilon$ (in our simulation $\varepsilon \in [0.0; 0.5]$) solitons may exist if $|\Delta| < \Delta_{cr}$. Thus, platicon repetition rate tuning range is $2\Delta_{cr}$. The critical value of modulation frequency mismatch grows almost linearly with the modulation depth value [see Fig. 6, right panel]. In comparison with bright solitons, repetition rate tuning range for platicons is significantly wider [see Fig. 7] for the same absolute values of the parameters. Moreover, at rather high values of modulation depth symmetry breaking occurs for bright solitons. It leads to the reduction of the tuning band since before symmetry breaking it is possible to decrease or increase the repetition rate equally. After symmetry breaking (at $\varepsilon \approx 0.29$ in Fig. 7) one tuning direction is suppressed and one can either increase or decrease repetition rate depending on the particular symmetry breaking case [compare lines marked with circles and triangles in Fig. 7].

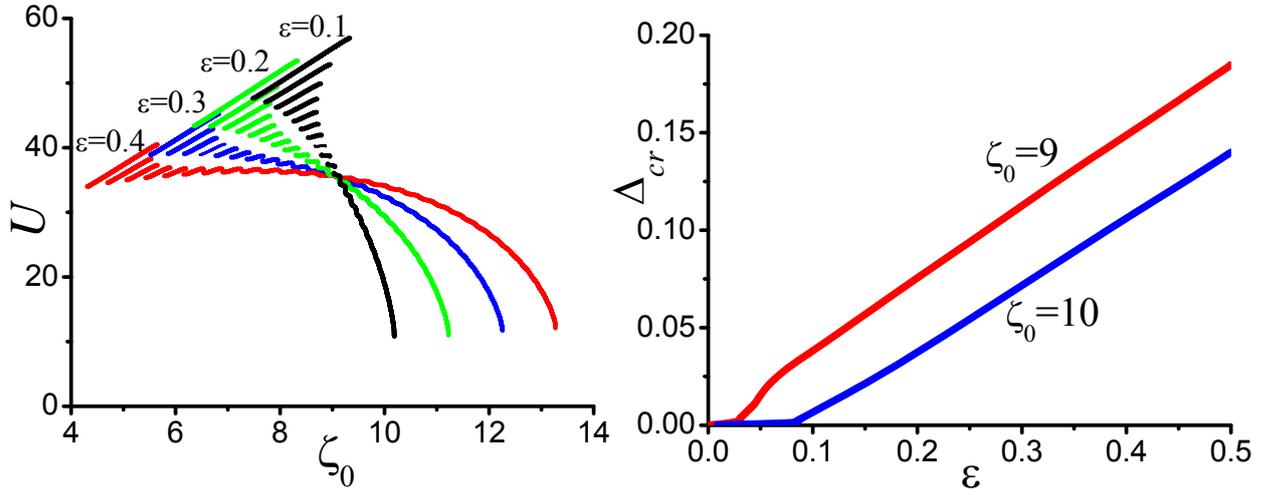

**Fig. 6.** (left panel) Platicon existence domains for different values of the modulation depth $\varepsilon$ at $\Delta = 0$. $U = \int_{-\pi}^{\pi} |\psi|^2 d\varphi$. (right panel) Modulation frequency mismatch critical value vs. modulation depth. All quantities are plotted in dimensionless units.

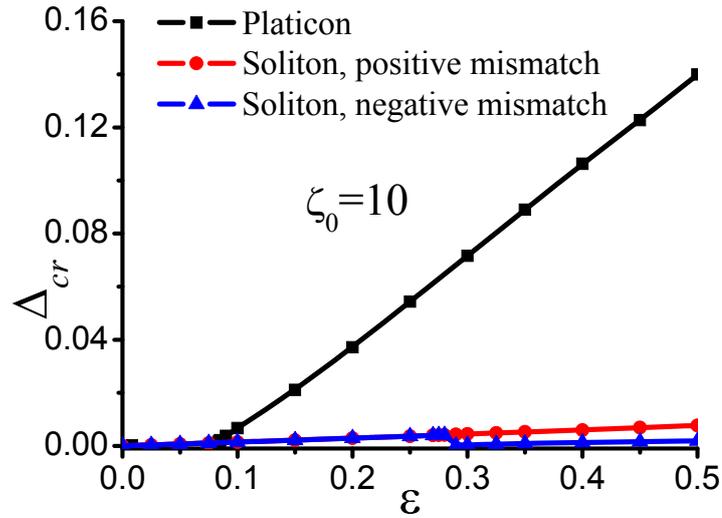

**Fig. 7.** Modulation frequency mismatch critical value vs. modulation depth for bright soliton and platicon for the same absolute values of the parameters ($F = 4$, $|\beta_2| = 0.02$). All quantities are plotted in dimensionless units.

If modulation is resonant the platicons resides at pump maximum (note, that stable bright solitons reside at pump minimum). In the presence of the modulation frequency mismatch platicons are shifted from pump maximum to the position where pump gradient can compensate drift term [see Fig. 8, left panel]. Shift direction is defined by the sign of the modulation frequency mismatch. Shift value increases with the growth of $\Delta$ and platicon spectrum becomes asymmetric [see Fig. 8, right panel]. If $|\Delta| > \Delta_{cr}$ platicons generated at $\Delta = 0$ decay upon propagation.

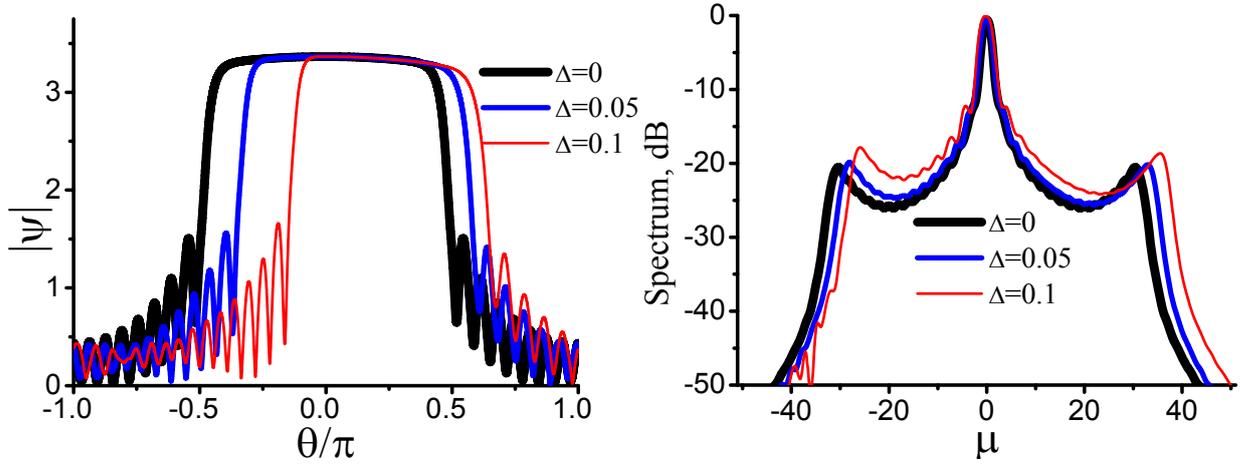

**Fig. 8.** Platicon profiles in dimensionless units (left panel) and spectra in dB (right panel) at $\zeta_0 = 10$, $\varepsilon = 0.4$ for different values of the modulation frequency mismatch $\Delta$.

$\Delta_{cr}$ also depends nonmonotonically on the detuning value $\zeta_0$ [see Fig. 9, left panel]. However, in the greater part of the existence domain, it decreases with the growth of $\zeta_0$. Note, that for detuning values providing several platicon solutions [see left panel in Fig. 6] we consider maximal possible value of modulation frequency mismatch in Fig. 9.

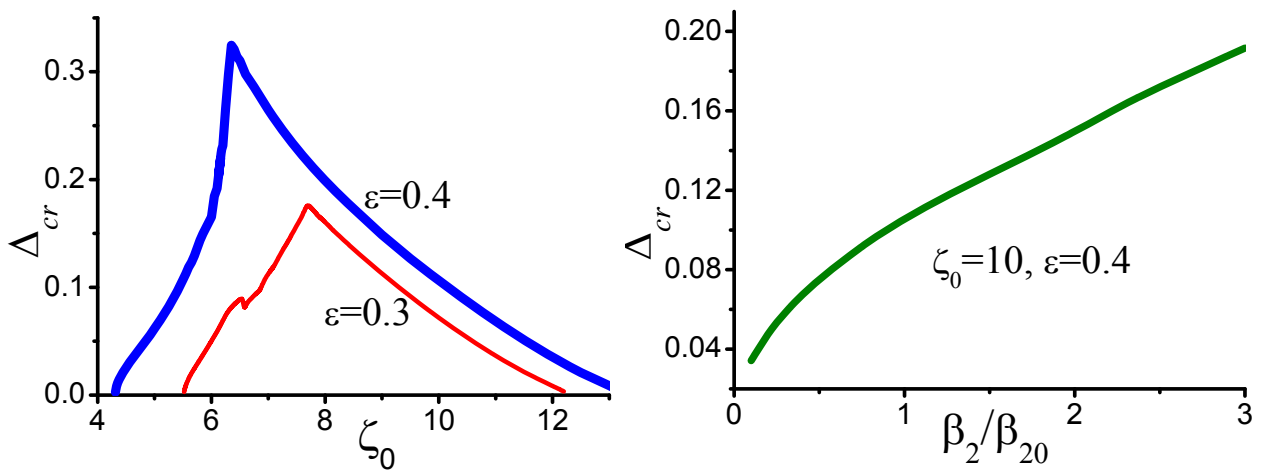

**Fig. 9.** (left panel) Mismatch critical value $\Delta_{cr}$ vs. pump detuning $\zeta_0$ for different values of $\varepsilon$. (right panel) Mismatch critical value $\Delta_{cr}$ vs. normalized GVD coefficient ($\beta_{20} = 0.02$). All quantities are plotted in dimensionless units.

We also found that $\Delta_{cr}$ increases with the growth of the GVD coefficient $\beta_2$. However, while for bright solitons this dependence was found to be linear, for platicons it not so [see Fig. 9, right panel].

Then, we introduced TOD term into Eq. 3 and using a platicon solution at $\Delta = 0$ and $\beta_3 = 0$ as an input we searched detuning range $\Delta_{cr-} \leq \Delta \leq \Delta_{cr+}$ providing the existence of stationary platicon-like solutions for different values of the third-order dispersion coefficient $\beta_3$. It was found out that TOD also affects platicon repetition rate tuning range [see Fig. 10]. Depending on its sign and value, TOD may strengthen or weaken the action of the drift term defined by the mismatch value. In the first case it leads to the decrease of the maximal mismatch value, in the opposite case – to the increase of $\Delta_{cr}$. So, at $\beta_3 = 0$ tuning range is symmetric with regard to the point $\Delta = 0$. At small values of $\beta_3$ tuning range becomes asymmetric and with further increase of $\beta_3$ it becomes shifted to the positive or negative values depending of the TOD sign. Thus, at significant values of $\beta_3$ platicons can exist only if modulation is nonresonant ($\Delta \neq 0$) since such parameters provides balance between modulation frequency mismatch and TOD. However, at large values of TOD platicon profile becomes indented significantly [see Fig. 11].

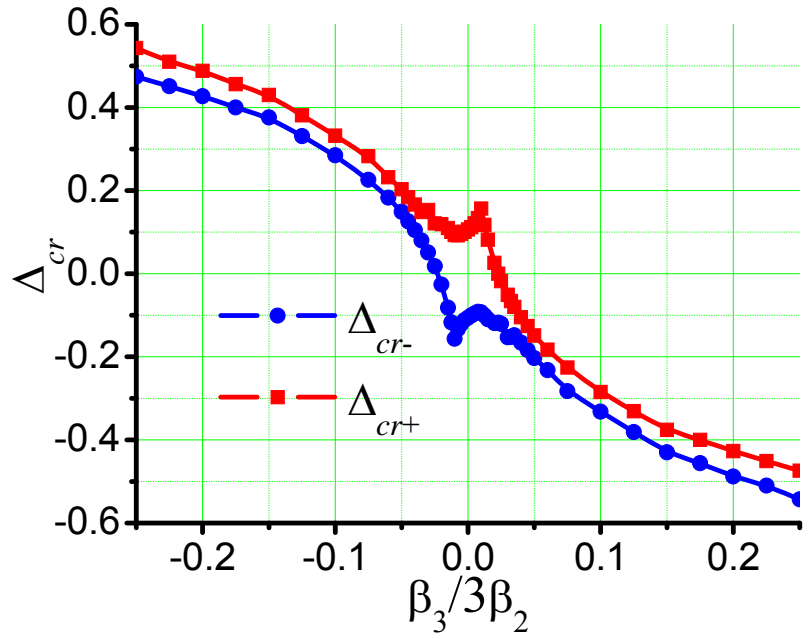

Fig. 10. Mismatch critical value vs. TOD coefficient at $\beta_2 = 0.02$, $\zeta_0 = 10$, $\varepsilon = 0.4$. All quantities are plotted in dimensionless units.

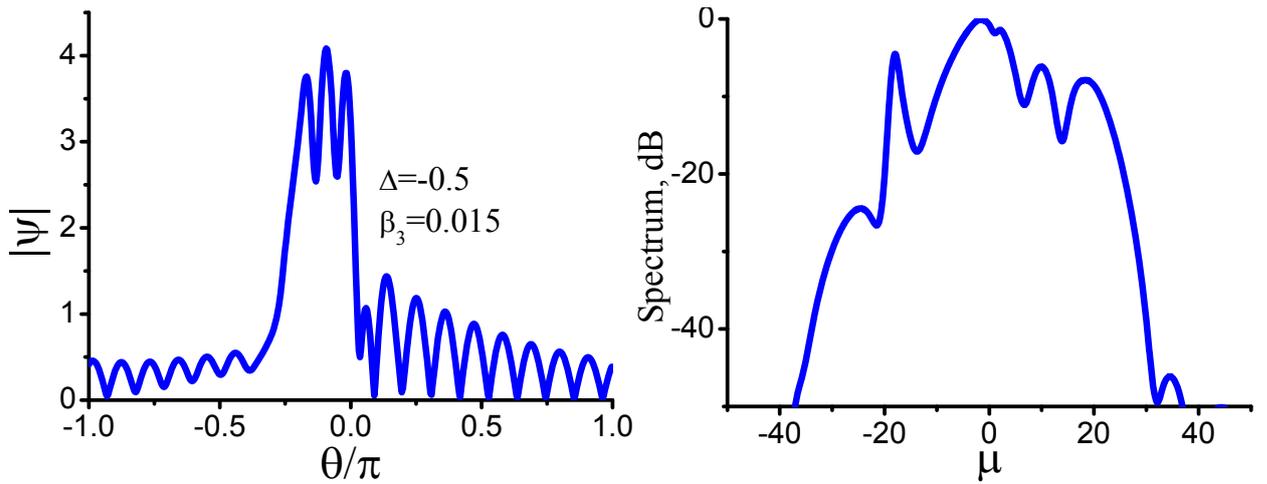

**Fig. 11.** Platicon profile in dimensionless units (left panel) and spectrum in dB (right panel) at $\zeta_0 = 10$, $\beta_2 = 0.02$, $\varepsilon = 0.4$, $\beta_3 = 0.015$, $\Delta = -0.5$.

We also revealed that while for repetition rate tuning of bright solitons one may use both AM and PM or their combination, for platicons it is possible with AM only. However, it can be expected that similar spectral purification mechanism of the external microwave signal frequency, leading to reduced phase noise of the output signal, reported for bright solitons [51] may exist for platicons.

### 5. Conclusion.

In conclusion, we found out that in the case of large values of FSR it is possible to use subharmonic phase modulation for platicon generation and analyzed PM parameters suitable for platicon generation that is useful for experimental realization. We also revealed the complex dynamics of platicons in the presence of TOD. It was shown that platicon dynamics depends on both modulation parameters and TOD coefficient value and it may be quite different from dynamics of dissipative Kerr solitons at anomalous GVD and dynamics of platicons generated due to pump mode shift. The obtained results are important for understanding the impact of different factors on the complex dynamics of nonlinear localized structures in optical microresonators. It was also demonstrated that it is possible to control the repetition rate of platicons by tuning modulation frequency and tuning range for platicons is much wider than for bright solitons. This effect may be useful for a variety of practical applications such as metrology, spectroscopy, and spectrometer calibration. TOD was shown to affect platicon repetition rate tuning range and may shift it away from the resonant modulation frequency. Introducing modulation frequency mismatch, it is possible to provide platicon existence at significant TOD values.

**Acknowledgement.**

This work was supported by the Russian Science Foundation (Project 17-12-01413).

**References**

1. T.J. Kippenberg, A.L. Gaeta, M. Lipson, M.L. Gorodetsky. Dissipative Kerr solitons in optical microresonators. Science. **361(6402)**, eaan8083 (2018).

2. T. Herr, V. Brasch, J. Jost, C. Wang, N. Kondratiev, M. Gorodetsky, and T. Kippenberg. Temporal solitons in optical microresonators. Nat. Photonics **8**, 145 (2014).

3. M.-G. Suh, Q.-F. Yang, K. Y. Yang, X. Yi, K. J. Vahala. Microresonator soliton dual-comb spectroscopy. Science **354**, 600–603 (2016).

4. N.G. Pavlov, G. Lihachev, S. Koptyaev, E. Lucas, M. Karpov, N.M. Kondratiev, I.A. Bilenko, T.J. Kippenberg, and M.L. Gorodetsky. Soliton dual frequency combs in crystalline microresonators. Opt. Lett. **42**, 514–517 (2017).

5. A. Dutt, C. Joshi, X. Ji, J. Cardenas, Y. Okawachi, K. Luke, A.L. Gaeta, M. Lipson. On-chip dual-comb source for spectroscopy. Science Advances. **4**, no. 3, e1701858 (2018).

6. M.-G. Suh, X. Yi, Y.-H. Lai, S. Leifer, I.S. Grudinin, G. Vasisht, E.C. Martin, M.P. Fitzgerald, G. Doppmann, J. Wang, D. Mawet, S.B. Papp, S.A. Diddams, C. Beichman, K. Vahala. Searching for Exoplanets Using a Microresonator Astrocomb. Nature Photonics. **13**, 25–30 (2019).

7. E. Obrzud, M. Rainer, A. Harutyunyan, M.H. Anderson, M. Geiselmann, B. Chazelas, S. Kundermann, S. Lecomte, M. Cecconi, A. Ghedina, E. Molinari, F.

Pepe, F. Wildi, F. Bouchy, T.J. Kippenberg, T. Herr. A microphotonic astrocomb. Nature Photonics. **13**, 31–35 (2019).

8. P. Trocha, D. Ganin, M. Karpov, M.H.P. Pfeiffer, A. Kordts, J. Krockenberger, S. Wolf, P. Marin-Palomo, C. Weimann, S. Randel, W. Freude, T.J. Kippenberg, C. Koos. Ultrafast optical ranging using microresonator soliton frequency combs. Science. **359**, 887–891 (2018).

9. S.B. Papp, K. Beha, P. Del'Haye, F. Quinlan, H. Lee, K.J. Vahala, and S.A. Diddams. Microresonator frequency comb optical clock. Optica **1**, 10 (2014).

10. P. Marin-Palomo, J.N. Kemal, M. Karpov, A. Kordts, J. Pfeifle, M.H.P. Pfeiffer, P. Trocha, S. Wolf, V. Brasch, M.H. Anderson, R. Rosenberger, K. Vijayan, W. Freude, T.J. Kippenberg, C. Koos. Microresonator-based solitons for massively parallel coherent optical communications. Nature, **546**, 274-279 (2017).

11. P. Liao, C. Bao, A. Kordts, M. Karpov, M.H.P. Pfeiffer, L. Zhang, Y. Cao, A. Almaiman, A. Mohajerin-Ariaei, F. Alishahi, A. Fallahpour, K. Zou, M. Tur, T.J. Kippenberg, and A.E. Willner. Effects of erbium-doped fiber amplifier induced pump noise on soliton Kerr frequency combs for 64-quadrature amplitude modulation transmission. Opt. Lett. **43**, 2495-2498 (2018).

12. X. Xue, M. Qi, and A.M. Weiner. Normal-dispersion microresonator Kerr frequency combs. Nanophoton. **5**, 244 (2016).

13. A.B. Matsko, A.A. Savchenkov, and L. Maleki. Normal group-velocity dispersion Kerr frequency comb. Opt. Lett. **37**, 43-45 (2012).

14. M.A. Foster, A.C. Turner, J.E. Sharping, B.S. Schmidt, M. Lipson and Alexander L. Gaeta. Broad-band optical parametric gain on a silicon photonic chip. Nature **441**, 960–963 (2006).

15. J.S. Levy, A. Gondarenko, M.A. Foster, A.C. Turner-Foster, A.L. Gaeta and M. Lipson. CMOS-compatible multiple-wavelength oscillator for on-chip optical interconnects. Nat. Photonics **4**, 37–40 (2010).

16. S.-W. Huang, H. Liu, J. Yang, M. Yu, D.-L. Kwong and C.W. Wong. Smooth and flat phase-locked Kerr frequency comb generation by higher order mode suppression. Sci. Rep. **6**, 26255 (2016).

17. W. Liang, A.A. Savchenkov, V.S. Ilchenko, D. Eliyahu, D. Seidel, A.B. Matsko, and L. Maleki. Generation of a coherent near-infrared Kerr frequency comb in a monolithic microresonator with normal GVD. Opt. Lett. **39**, 2920–2923 (2014).

18. X. Xue, Y. Xuan, Y. Liu, P.-H. Wang, S. Chen, J. Wang, D.E. Leaird, M. Qi and A.M. Weiner. Mode-locked dark pulse Kerr combs in normal-dispersion microresonators. Nat. Photonics **9**, 594–600 (2015).

19. V.E. Lobanov, G. Lihachev, T. J. Kippenberg, and M.L. Gorodetsky. Frequency combs and platicons in optical microresonators with normal GVD. Opt. Express **23**, 7713 (2015).

20. Y. Liu, et al. Investigation of mode coupling in normal-dispersion silicon nitride microresonators for Kerr frequency comb generation. Optica. **1**, 137-144 (2014).

21. X. X. Xue, Y. Xuan, P.-H. Wang, Y. Liu, D.E. Leaird, M. Qi, A.M. Weiner. Normal-dispersion microcombs enabled by controllable mode interactions. Laser Photonics Rev. **9**, L23–L28 (2015).

22. N. M. Kondratiev, V. E. Lobanov, A. V. Cherenkov, A. S. Voloshin, N. G. Pavlov, S. Koptyaev, and M. L. Gorodetsky. Self-injection locking of a laser diode to a high-Q WGM microresonator. Opt. Express **25**, 28167-28178 (2017)

23. P. Parra-Rivas, E. Knobloch, D. Gomila, L. Gelens. Dark solitons in the Lugiato-Lefever equation with normal dispersion. Phys. Rev. A **93**, 063839 (2016).

24. P. Parra-Rivas, D. Gomila, E. Knobloch, S. Coen, and L. Gelens. Origin and stability of dark pulse Kerr combs in normal dispersion resonators. Opt. Lett. **41**, 2402-2405 (2016).

25. N.N. Rosanov and G.V. Khodova. Diffractive autosolitons in nonlinear interferometers. J. Opt. Soc. Am. B 7, 1057-1065 (1990).

26. X. Xue, P.-H. Wang, Y. Xuan, M. Qi, A.M. Weiner. Microresonator Kerr frequency combs with high conversion efficiency. Las. Photon. Rev. **11**, 1600276 (2017).

27. C. Bao, L. Zhang, A. Matsko, Y. Yan, Z. Zhao, G. Xie, A.M. Agarwal, L.C. Kimerling, J. Michel, L. Maleki, and A.E. Willner. Nonlinear conversion efficiency in Kerr frequency comb generation. Opt. Lett. **39**, 6126–6129 (2014).

28. P.-H. Wang, J.A. Jaramillo-Villegas, Y. Xuan, X. Xue, C. Bao, D.E. Leaird, M. Qi, and A.M. Weiner. Intracavity characterization of micro-comb generation in the single-soliton regime. Opt. Express **24**, 10890–10897 (2016).

29. V.E. Lobanov, A.V. Cherenkov, A.E. Shitikov, I.A. Bilenko, M.L. Gorodetsky, Dynamics of platicons due to third-order dispersion. Eur. Phys. J. D **71**, 185 (2017).

30. A.V. Cherenkov, V.E. Lobanov, and M.L. Gorodetsky. Dissipative Kerr solitons and Cherenkov radiation in optical microresonators with third-order dispersion. Phys. Rev. A. **95**, 033810 (2017).

31. P. Parra-Rivas, D. Gomila, L. Gelens. Coexistence of stable dark- and bright-soliton Kerr combs in normal-dispersion resonators. Phys. Rev. A **95**, 053863 (2017).

32. A.V. Cherenkov, N.M. Kondratiev, V.E. Lobanov, A.E. Shitikov, D.V. Skryabin, and M.L. Gorodetsky. Raman-Kerr frequency combs in microresonators with normal dispersion. Opt. Express **25**, 31148-31158 (2017).

33. V.E. Lobanov, G. Lihachev, and M.L. Gorodetsky. Generation of platicons and frequency combs in optical microresonators with normal GVD by modulated pump. EPL. **112**, 54008 (2015).

34. H. Liu, S.-W. Huang, J. Yang, M. Yu, D.-. Kwong, and C. W. Wong. "Bright square pulse generation by pump modulation in a normal GVD microresonator," in Conference on Lasers and Electro-Optics, OSA Technical Digest (online) (Optical Society of America, 2017), paper FTu3D.3 (2017).

35. V.E. Lobanov, G.V. Lihachev, N.G. Pavlov, A.V. Cherenkov, T.J. Kippenberg, and M.L. Gorodetsky. Harmonization of chaos into a soliton in Kerr frequency combs. Opt. Express **24**, 27382-27394 (2016).

36. D.C. Cole, J.R. Stone, M. Erkintalo, K.Y. Yang, X. Yi, K.J. Vahala, and S.B. Papp. Kerr-microresonator solitons from a chirped background. Optica **5**, 1304-1310 (2018).

37. Y. K. Chembo and N. Yu. Modal expansion approach to optical-frequency-comb generation with monolithic whispering-gallery-mode resonators. Phys. Rev. A **82**, 033801 (2010).

38. L.A. Lugiato and R. Lefever. Spatial dissipative structures in passive optical systems. Phys. Rev. Lett. **58**, 2209 (1987).

39. Y.K. Chembo and C.R. Menyuk. Spatiotemporal Lugiato-Lefever formalism for Kerr-comb generation in whispering-gallery-mode resonators. Phys. Rev. A. **87**, 053852 (2013).

40. S. Coen, H. G. Randle, T. Sylvestre, and M. Erkintalo. Modeling of octave-spanning Kerr frequency combs using a generalized mean-field Lugiato-Lefever model. Opt. Lett. **38(1)**, 37 (2013).

41. V. Brasch, M. Geiselmann, T. Herr, G. Lihachev, M H.P. Pfeifer, M.L. Gorodetsky, and T.J. Kippenberg. Photonic chip-based optical frequency comb using soliton Cherenkov radiation. Science. 351(6271), 357 (2016).

42. S. Wang, H. Guo, X. Bai, and X. Zeng. Broadband Kerr frequency combs and intracavity soliton dynamics influenced by high-order cavity dispersion. Opt. Lett. **39(10)**, 2880 (2014).

43. C. Milián, A. V. Gorbach, M. Taki, A. V. Yulin, and D. V. Skryabin. Solitons and frequency combs in silica microring resonators: Interplay of the Raman and higher-order dispersion effects. Phys. Rev. A **92**, 033851 (2015).

44. C. Milián and D.V. Skryabin. Soliton families and resonant radiation in a micro-ring resonator near zero group velocity dispersion. Opt. Exp. **22(3)**, 3732 (2014).

45. D.V. Skryabin and Y.V. Kartashov. Self-locking of the frequency comb repetition rate in microring resonators with higher order dispersions. Opt. Express **25**, 27442 (2017).

46. Y. He, S. Wang, X. Zeng. Dynamics of Dispersive Wave Emission from Dark Solitons in Kerr Frequency Combs. IEEE Phot. Journ. **8(6)**, 7102508 (2016).

47. T. Hansson, D. Modotto, and S. Wabnitz. On the numerical simulation of Kerr frequency combs using coupled mode equations. Opt. Commun. **312**, 134 (2014).

48. B. Razavi. A study of injection locking and pulling in oscillators. IEEE J. Solid-State Circuits. **39**, 1415 (2004).

49. J. K. Jang, M. Erkintalo, S. Coen, and S. G. Murdoch. Temporal tweezing of light through the trapping and manipulation of temporal cavity solitons. Nat. Commun. **6**, 7370 (2015).

50. H. Taheri, A. B. Matsko, and L. Maleki. Optical lattice trap for Kerr solitons Eur. Phys. J. D. **71**, 153 (2017).

51. W. Weng, E. Lucas, G. Lihachev, V.E. Lobanov, H. Guo, M.L. Gorodetsky, T.J. Kippenberg. Spectral purification of microwave signals with disciplined dissipative Kerr solitons. Phys. Rev. Lett. **122 (1)**, 013902 (2019).

52. I. Hendry, W. Chen, Y. Wang, B. Garbin, J. Javaloyes, G.-L. Oppo, S. Coen, S. G. Murdoch, M. Erkintalo. Spontaneous symmetry breaking and trapping of temporal Kerr cavity solitons by pulsed or amplitude-modulated driving fields. Phys. Rev. A **97**, 053834 (2018).